\documentstyle{article}
\begin{document}
\newcommand{\C}{{\cal C}}
\begin{flushright}
{\sf Portsmouth University\\
Relativity and Cosmology Group\\
{\em Preprint} RCG 96/1}
\end{flushright}
\[ \]
{\Large\bf Anisotropic Observations in Universes with Nonlinear 
Inhomogeneity}
\[ \]
Neil P Humphreys${}^{\dag}$,
Roy Maartens${}^{\dag}{}^{\star}$, and
David R Matravers${}^{\dag}$
\[ \]
${}^{\dag}${\footnotesize School of Mathematical Studies, 
Portsmouth
University, Portsmouth PO1 2EG, England}\\
${}^{\star}${\footnotesize Member of Center for Nonlinear Studies,
Witwatersrand University, 2050 South Africa}
\[ \]
{\bf Abstract. }
We calculate the off--center observational relations in a 
spherically symmetric
dust universe that is inhomogeneous at small redshifts. In 
contrast to the  usual model, in which the {\sc CMBR} dipole is 
interpreted as a Doppler 
effect due to peculiar velocity, our model explores an 
alternative interpretation, in 
which the {\sc CMBR} dipole is non--Doppler, and the observer is 
comoving with the mean matter flow. We do not assume a background 
frame relative to which peculiar velocities are calculated. 
Our analysis is fully nonlinear and the density contrast is not 
assumed to be small.
We obtain exact expressions for the Hubble and deceleration 
parameters, and find that both parameters have quadrupole 
anisotropies, but no dipoles. A simple numerical 
procedure for calculating the {\sc CMBR} dipole anisotropy 
in our model is presented, and the observed $0.1$\% dipole is 
shown to be reproducible with a reasonable choices of parameters.

\section*{1. Introduction}
Cosmological models with inhomogeneous spherically symmetric 
regions have
been extensively studied (e.g. Wesson 1979; Raine and Thomas 1981; 
Lynden--Bell {\em et al}. 1988; Bertschinger and Juszkiewicz 1988; 
Bertschinger and Dekel 1989; Panek 1992; M\'{e}sz\'{a}ros 1994;
Moffat and Tatarski 1995; Tomita 1995). With 
some exceptions (e.g. Tomita and Watanabe 1989; Paczynski and 
Piran 1990; Arnau {\em et al}. 1994), 
the inhomogeneity in the matter is generally 
treated as a perturbation of a homogeneous background, which 
restricts the analysis to small density contrast and small 
peculiar gravitational potential in a linearised theory.
The {\sc CMBR} dipole is then interpreted as a Doppler effect 
arising from our peculiar motion relative to the background.

Linearised inhomogeneous models with a position--dependent Hubble
parameter $H_{0}$ can be used to explain why
measurements of $H_{0}$ (see Riess {\em et al}. (1995) and Van den 
Bergh (1995) for recent results) tend to give relatively high 
values from methods dependent on local observations, and lower 
values from methods
based on more distant observations. (The Hubble parameter 
relevant to large--scale cosmology should be low to avoid 
contradiction between the 
universal age calculated from $H_{0}$ and $\Omega_{0}$, and the
limits due to the age of constituents of the universe).

In this paper the inhomogeneity is treated self--consistently, 
without 
linearised perturbations, thus allowing for nonlinear peculiar 
gravitational potential and density contrast (while also 
incorporating the case of small density perturbations). We 
therefore consider cosmological models which may be homogeneous 
on large scales, but 
which can incorporate arbitrary isotropic inhomogeneity on 
smaller scales (say about $100\;$Mpc). In order to exploit
the possibilities of such models, while simultaneously exploring 
alternatives to the standard models, we take the dipole 
anisotropy of the {\sc CMBR}, and any anisotropies in the Hubble 
and deceleration parameters, to arise not from the observer's 
peculiar motion (relative to some 
background), but from the observer's off--center location, i.e. 
from the inhomogeneous gravitational field. Galaxies follow 
unperturbed motion in this field, so from their viewpoint there 
is no peculiar velocity. 
For comoving observers away from the center of symmetry, 
the {\sc CMBR} dipole anisotropy will be interpreted as a 
cosmological effect, rather than the effect of local peculiar 
motion. A similar model of the
{\sc CMBR} dipole was given by Paczynski and Piran (1990), and 
non--Doppler dipoles arising from large--scale perturbations were 
considered by Bildhauer and Futamase (1991), and Langlois and 
Piran (1995). The {\sc CMBR} itself
cannot distinguish between these two interpretations. Indeed, 
the observer's peculiar velocity is not directly observed, but 
postulated as a reasonable source of the {\sc CMBR} dipole. Our 
model thus investigates an alternative explanation.

As pointed out by Paczynski and Piran (1990), observations 
have not decisively established the reality of large--scale 
departures from a mean
Hubble flow, so that alternative models of the {\sc CMBR} 
dipole can help to
develop new ways of testing the standard model. It turns out 
that both $H_{0}$ and $q_{0}$ in our model (i.e. a freely 
falling off--center observer) have \em no \rm dipole 
anisotropy, whereas the standard model (i.e. peculiar motion 
towards a Great--Attractor--like structure) does predict dipoles 
in $H_{0}$ and $q_{0}$ ({\em cf}. Nakao {\em et al}. 1995). Thus, in 
principle, measurements of $H_{0}$ via the 
distance--redshift observations
could distinguish between these alternatives. 

Exact formulas for the Hubble and deceleration 
parameters are given in \S 4. 
These arise from the local solution of the null--geodesic 
equation near the  off--center observer, as discussed in \S 2.
In \S 5 we show how to calculate the {\sc CMBR} dipole, and in 
the appendix the observed value is obtained using a reasonable 
choice of cosmological parameters.

\section*{2. Off--Center Geometry}
The Lema\^{\i}tre--Tolman--Bondi ({\sc LTB}) metric (Bondi 1947) 
gives the geometry corresponding to a spherically symmetric 
space--time with a pressure--free dust matter distribution, 
prior to the formation of shell--crossing singularities:
\begin{equation}
ds^{2}=-dt^{2}+\frac{R^{\prime}(r,t)^2}{1+f(r)}dr^2+R(r,t)^2 
(d\theta^2+\sin^2\theta\;d\phi^2)  \label{metric}
\end{equation}
where $x^{0}\equiv t$ is cosmic proper time and $x^{i}\equiv
\left\{r,\theta,\phi\right\}$ are comoving. The arbitrary 
function $f$ determines the curvature of the spatial 
hypersurfaces. The units are chosen such that $c=G=1$, and the 
cosmological constant $\Lambda=0$. 
Einstein's field equations reduce to
\begin{eqnarray}
&&\dot{R}^{2}=\frac{2M(r)}{R}+f\label{feqn}\\
&&\rho=\frac{M^{\prime}}{4\pi R^{\prime}R^{2}} 
\label{density}
\end{eqnarray}
where $\rho$ is the proper density, and $M(r)$ is the
gravitational mass inside the sphere of radius $r$. A prime 
denotes $\partial/\partial r$, and a dot $\partial/\partial t$. 
The four--velocity of the dust is $u^{\mu}=\delta^{\mu}_{0}$.
 
Let $P_{0}$ label the  
spacetime event of an off--center observer at time $t=t_{0}$, 
i.e. the event $\{t=t_{0},r=r_{0},\theta=\pi/2,\phi=0\}$.

With $k^{\mu}$ the tangent to the geodesics 
$x^{\mu}(\upsilon)$, the geodesic 
equations $k^{\mu}_{\;\; ;\nu}k^{\nu}=0$ take the form
\begin{eqnarray}
& &\frac{d^2 t}{d\upsilon^2}+\frac{\dot{R}^{\prime}
R^{\prime}}{1+f}\left(\frac{dr}{d\upsilon}\right)^2
+\dot{R}RL^{2}=0
\label{geo1}\\
& &\frac{d^2 r}{d\upsilon^2}+2\frac{\dot{R}^{\prime}}
{R^{\prime}}\frac{dt}{d\upsilon}\frac{dr}{d\upsilon}+\left(
\frac{R^{\prime\prime}}{R^{\prime}}-\frac{f^{\prime}}
{2(1+f)}\right)\left(\frac{dr}{d\upsilon}\right)^2
-(1+f)\frac{R}{R^{\prime}}L^{2}=0 
\label{geo2}\\
& &\frac{d^2\theta}{d\upsilon^2}+2\frac{\dot{R}}{R}
\frac{d\theta}{d\upsilon}\frac{dt}{d\upsilon}
+2\frac{R^{\prime}}{R}\frac{dr}{d\upsilon}\frac{d\theta}
{d\upsilon}-\sin{\theta}\cos{\theta}\left(
\frac{d\phi}{d\upsilon}\right)^2=0 
\label{geo3}\\
& &\frac{d^2\phi}{d\upsilon^2}+2\frac{\dot{R}}{R}\frac{d\phi}
{d\upsilon}\frac{dt}{d\upsilon}+2\frac{R^{\prime}}{R}\frac{dr}
{d\upsilon}\frac{d\phi}{d\upsilon}+2\cot{\theta}
\frac{d\theta}{d\upsilon}\frac{d\phi}{d\upsilon}=0 \label{geo4}
\end{eqnarray}
where $L^{2}\equiv (d\theta/d\upsilon)^2+\sin^2\theta 
(d\phi/d\upsilon)^2$. The first integral
\begin{equation}
-\left(\frac{dt}{d\upsilon}\right)^2+\frac{R^{\prime 2}}{1+f}
\left(\frac{dr}{d\upsilon}\right)^2+R^2 L^{2}=0  \label{g5}
\end{equation}
specifies that the geodesics are null. If the choice 
$\upsilon=0$ at $P_{0}$ is made, then  
the past null--geodesics through $P_{0}$ are locally expressible 
as
\begin{equation}
x^{\alpha}(\upsilon)=\sum_{n=0}^{3}\left(x^{\alpha}\right)_{n}
\upsilon^{n}\;+O(\upsilon^{4})
\label{taylor}
\end{equation}
and the functions $R$ and $f$ may similarly be assumed
Taylor--expandable. 
Throughout the paper a subscript `$0$' denotes 
evaluation at $P_{0}$. The geodesic equations 
(\ref{geo1}--\ref{geo4}) may be 
solved locally by equating coefficients of powers of $\upsilon$. 
The solution is of the form $\left\{r_{2},r_{3},\cdots,\right.$ 
$t_{2},t_{3},\cdots,$ $\theta_{2},\theta_{3},\cdots,$ $\left.
\phi_{2},\phi_{3}\cdots\right\}$ with each coefficient a function 
of the observational angles $\vartheta$ and $\varphi$ at $P_{0}$.

Since affine parameters remain affine under linear 
transformations, $\upsilon$ may be fixed by choosing
\begin{equation}
t_{1}=-1      \label{gauge}
\end{equation} 
This choice forces $\upsilon$ to increase down 
geodesics of incoming light, and is equivalent to setting 
$u_{\mu}k^{\mu}=1$ at $P_{0}$ (Ellis {\em et al}. 1985). 
Evaluation 
of equation (\ref{g5}) to lowest order in $\upsilon$ results in
\begin{equation}
r_{1}=\left(-1\right)^{\epsilon}\left[\frac{\sqrt{1+f}}
{R^{\prime}}\sqrt{1-R^2 (\theta_{1}^2+\phi_{1}^2)}\right]_{0}
\label{r1}\end{equation}
The factor $\epsilon=0$ (or $1$) is determined by the motion 
of rays at $P_{0}$ toward (or away) from the center 
of symmetry.

Equations (\ref{g5}) and (\ref{r1}) are equivalent, since 
geodesics that are initially null  
must remain null. The four components of $k^{\mu}$ 
at $P_{0}$ may be interpreted as follows: $t_{1}$ fixes the 
gauge of the affine parameter by equation (\ref{gauge}), 
$r_{1}$ specifies the geodesics are null by equation (\ref{r1}),
and $\theta_{1}$ and $\phi_{1}$ determine 
the 2--sphere of observer angles $\left\{\vartheta,\varphi\right\}$ 
at $P_{0}$. 

Calculation of $\vartheta$ and $\varphi$ in terms of $\theta_{1}$, 
$\phi_{1}$, and $\epsilon$, is as follows. Let $I^{\mu}$, 
$J^{\mu}$, 
and $K^{\mu}$ be unit vectors in the surface 
$\left\{t=t_{0}\right\}$, with $I^{\mu}$ pointing radially outward 
from the center of symmetry, $K^{\mu}$ pointing in the direction  
associated with $-{\partial}/\partial\theta$, and $J^{\mu}$ 
pointing in the direction of an observed light ray:
\begin{eqnarray}
& &I^{\mu}=\left[\frac{\sqrt{1+f}}{R^{\prime}}\delta^{\mu}_{r} 
\right]_{0}\\
& &J^{\mu}=\left(0,\;\;\;\frac{\sqrt{ (1+f)\left[1-R^{2}
(\theta_{1}^{2}
+\phi_{1}^{2})\right] }}{R^{\prime}},\;\;\;\theta_{1},\;\;\;
\phi_{1}\right)_{0} \label{j}\\ 
& &K^{\mu}=-\left[\frac{1}{R}\delta^{\mu}_{\theta}\right]_{0}
\end{eqnarray}
Let $\vartheta$ and $\varphi$ be defined (relative to $\theta$, 
$\phi$, and $\epsilon$) via 
\begin{equation}
\epsilon=0=\phi_{1}\Leftrightarrow \varphi=0,\;\;\;\;\;
\theta_{1}=0\Leftrightarrow\vartheta=\frac{\pi}{2} 
\end{equation}
Then $\varphi$ is given by $\cos{\varphi}=I^{\mu}J_{\mu}$ with 
$\theta_{1}=0$, which yields
\begin{equation}
\varphi=\epsilon\pi+\left(-1\right)^{\epsilon}\arcsin\left(R
\phi_{1}\right)_{0} \label{angle1}\end{equation}
Also $\cos{\vartheta}=J^{\mu}K_{\mu}$ implies
\begin{equation}
\vartheta=\arccos\left(-R\theta_{1}\right)_{0} \label{angle2}
\end{equation}
Equations (\ref{angle1}) and (\ref{angle2}) invert to
\begin{equation}
\phi_{1}=\left[\frac{\sin\varphi}{R}\right]_{0},\;\;\;\;\;\;\;\;
\theta_{1}=-\left[\frac{\cos\vartheta}{R}\right]_{0} 
\label{unangles}\end{equation}
Now, by spherical symmetry, 
just one angle is required to describe the off--center 
observations: the angle $\psi$ between $I^{\mu}$ and the 
direction of observation at $P_{0}$. Observations with common 
$\psi$ must be indistinguishable. Attention may therefore be 
restricted to the plane $\left\{\theta(\upsilon)=\pi/2=
\vartheta\right\}$, for which $\psi=\varphi$. 

\section*{3. Matching Constraints}
We require that the inhomogeneous models become homogeneous
on a sufficiently large distance scale. For simplicity it is 
supposed that the transition occurs on a well--defined comoving 
spherical boundary without any surface density layer (Israel 
1966). The coordinates $x^{\mu}$ may be extended beyond the 
boundary (at $r=\chi$, say) and into the external {\sc FLRW}
region. For $r<\chi$ the inhomogeneous {\sc LTB} solution is 
\begin{eqnarray}
f=0:\;\;&R=\left(9M/2\right)^{\frac{1}{3}}\left(t-\beta
\right)^{\frac{2}{3}}&\label{f=0inhom}\\
f>0:\;\;&R=M\left(\cosh\eta-1\right)/f,&\;\;
\sinh\eta-\eta=f^{\frac{3}{2}}\left(t-\beta\right)/M \label{f>0 i}
\\
0>f>-1:\;\;&R=M\left(\cos\eta-1\right)/f,&\;\;
\eta-\sin\eta=|f|^{\frac{3}{2}}\left(t-\beta\right)/M  \label{f<0 i}
\end{eqnarray}
in the parabolic, hyperbolic, and elliptic cases 
respectively. The function $\beta(r)$ gives the big--bang 
hypersurface. For $r>\chi$ the homogeneous {\sc LTB} 
({\sc FLRW}) solution is
\begin{eqnarray}
f=0:\;\;&R=\left(9Mt^{2}/2\right)^{\frac{1}{3}}&\label{flrw1}\\
f=\alpha\left(2M\right)^{\frac{2}{3}}>0:\;\;&R=\left(\cosh\eta 
-1\right)
\left(M/4\alpha^{3}\right)^{\frac{1}{3}},&\;\;
\sinh\eta-\eta=2\alpha^{\frac{3}{2}}t\label{flrw2}\\
0>f=-\alpha\left(2M\right)^{\frac{2}{3}}>-1:\;\;&R=\left(
1-\cos\eta\right)
\left(M/4\alpha^{3}\right)^{\frac{1}{3}},&\;\;
\eta-\sin\eta=2\alpha^{\frac{3}{2}}t\label{flrw3}
\end{eqnarray}
where $\alpha$ is a constant, and the large--scale {\sc FLRW} 
Hubble and density parameters are given by
\begin{eqnarray}
f=0:\;\;&{\cal H}=2/3t,&\;\;
\Omega=1\label{global1}\\
f>0:\;\;&{\cal H}=2\alpha^{\frac{3}{2}}\sinh\eta/\left(
\cosh\eta-1\right)^{2},&\;\;\Omega=2\left(\cosh\eta-1\right)/
\sinh^{2}\eta
\\
0>f>-1:\;\;&{\cal H}=2\alpha^{\frac{3}{2}}\sin\eta/\left(
1-\cos\eta\right)^{2},&\;\;\Omega=2\left(1-\cos\eta\right)
/\sin^{2}\eta\label{global}
\end{eqnarray}
The boundary hypersurface has intrinsic metric  
$K_{\mu\nu}=g_{\mu\nu}-n_{\mu}n_{\nu}$ and extrinsic curvature 
$n_{\mu;\nu}K^{\mu}_{\alpha}K^{\nu}_{\beta}$, where $n_{\mu}$ 
is the unit normal.
The Darmois junction  conditions (for the case with no surface 
layer at $r=\chi$) are (Israel 1966)
that the intrinsic metric and the extrinsic curvature, 
evaluated on the two sides of $r=\chi$, are equal.
Restricting attention to solutions which have the same 
characteristic curvatures (parabolic, hyperbolic, or elliptic) 
on both sides of $r=\chi$, the conditions are
\begin{eqnarray}
{\rm all }\;\;f&:\;\;&\lim_{\delta\rightarrow 0}\left[M(\chi+
\delta)-M(\chi-\delta)\right]=0\label{continuity}\\
f=0&:\;\;&\beta(\chi)=0\label{f=0con}\\
f>0&:\;\;&\beta(\chi)=0,\;\;\;\;\;\;f(\chi)=\alpha\left[
2M(\chi)\right]^{\frac{2}{3}}\\
0>f>-1&:\;\;&\beta(\chi)=0,\;\;\;\;\;\;f(\chi)=-\alpha\left[
2M(\chi)\right]^{\frac{2}{3}}
\end{eqnarray}
Papapetrou (1978) successfully matched an interior collapsing 
($f<0$) {\sc LTB} solution to the Einstein--de Sitter 
model 
(for which $f=0$). It may also be possible to obtain an 
internal {\sc LTB} solution consisting of several zones with 
different solution branches ($f<0,f=0,f>0$), obeying the Darmois 
conditions on their concentric spherical interfaces. These are 
however highly non--trivial models, in which the avoidance of
shell--crossing caustics is a difficult problem. (When dust shells 
cross the zero--pressure condition of {\sc LTB} is violated). 
An analysis of some of the issues relating to the matching of
{\sc LTB} solutions will be the subject of subsequent research.

\section*{4. Off--Center Observations: Galaxies}
In this paper, idealised cosmological observations are assumed. 
In particular, it is supposed that area distances (or 
equivalently luminosity distances) and number counts are known 
as functions of the redshift. Hence any statistical, averaging, 
smoothing, and distance--dependent selection effects are neglected. 
As is well known (Maartens {\em et al}. 1996) {\sc LTB} models  
are completely specified by complete measurements of the area 
distance and number count as functions of the redshift. In what 
follows, these two functions are expanded about $P_{0}$, up to the  
first two non--vanishing derivatives in the redshift. 

The redshift $z$ as measured at $P_{0}$ is given by 
$1+z=u^{\mu}k_{\mu}$ (Ellis {\em et al}. 1985), and by equation 
(\ref{gauge}):
\begin{equation}
z=-\frac{dt(\upsilon)}{d\upsilon}-1     \label{z}
\end{equation}
The observer area distance $D$ at $P_{0}$ may be obtained by 
considering an 
observational coordinate system (Ellis {\em et al}. 1985) 
at $P_{0}$. 
The observational coordinates are
$\tilde{x}^{\mu}\equiv\left\{w,\upsilon,\vartheta,
\varphi\right\}$, where $\left\{w=\right.$constant$
\left.\right\}$ are the past light cones from the observer's 
world--line at $P_{0}$.
The metric tensor in observational coordinates is then
\begin{equation}
\tilde{g}_{\mu\nu}=g_{\alpha\gamma}\frac{\partial x^{\alpha}}
{\partial \tilde{x}^{\mu}}\frac{\partial x^{\gamma}}
{\partial \tilde{x}^{\nu}}    \label{trans}
\end{equation}
Thus the angular part of the transformed metric  
$\tilde{g}_{\vartheta\vartheta},\tilde{g}_{\varphi\varphi},
\tilde{g}_{\vartheta\varphi}$ on the
past--light cone of $P_{0}$ can be calculated locally from 
the null--geodesic solution and equations (\ref{unangles}). 
Then $D$ follows from the formula (Ellis {\em et al}. 1985) 
\begin{equation}
D^{4}\sin^{2}{\vartheta}=\tilde{g}_{\vartheta\vartheta}
\tilde{g}_{\varphi\varphi}-
\left(\tilde{g}_{\vartheta\varphi}\right)^2 \label{D}
\end{equation}
Equation (\ref{D}) and the local null--geodesic solution lead to
\begin{equation}
\left(\frac{dD}{dz}\right)_{0}=\frac{1}{{\cal C}_{1}+
{\cal C}_{2}\cos^{2}\psi} \label{hresult}
\end{equation}
\begin{equation}
\left(\frac{d^{2}D}{dz^{2}}\right)_{0}=\frac{\C_{3}+
\C_{4}|\cos\psi|+\C_{5}\cos^{2}\psi+\C_{6}|\cos\psi|^{3}}
{\left({\cal C}_{1}+{\cal C}_{2}\cos^{2}\psi\right)^{3}}  
\label{decel}\end{equation}
where the constants $\C_{I}$ are given in the appendix. 

The Hubble parameter $H$ of a spherically symmetric 
inhomogeneous universe is often given two definitions (e.g. 
Moffat and Tatarski 1995):
\begin{equation}
H_{r}=\dot{R}/R,\;\;\;\;\;\;H_{\phi}=\dot{R}^{\prime}/R^{\prime} 
\label{moffat}
\end{equation}
which are to be interpreted as radial and azimuthal expansion 
rates (generally the expansion is anisotropic since the shear 
is non--vanishing). The two definitions agree at the center 
of symmetry, as follows
from equation (\ref{feqn}) and the 
behaviours of $f$, $M$, and $R$ near the center (Bondi 1947):
\begin{equation}
R(r,t)=R^{\prime}(0,t)r+O(r^{2}),\;\;\;\;\;\;
f(r)=\frac{1}{2}f^{\prime\prime}(0)r^{2}+O(r^{3}),\;\;\;\;\;\;
M(r)=\frac{1}{6}M^{\prime\prime\prime}(0)r^{3}+O(r^{4}) 
\end{equation}
Using equations (\ref{moffat}), we obtain
\begin{equation}
H_{\phi}\left(0,t\right)=R^{\prime}(0,t)^{-1}\sqrt{
M^{\prime\prime\prime}(0)/6R^{\prime}(0,t)+\frac{1}{2}
f^{\prime\prime}(0)}=H_{r}\left(0,t\right)
\end{equation}
Moffat 
and Tatarski (1995) have posed the question of which one is the 
analogue of the {\sc FLRW} $H_{0}$, when off--center.  We resolve 
this issue by the use of a covariant definition:
\begin{equation}
H_{0}=\frac{1}{\delta l}\frac{d\left(\delta l\right)}{dt} 
\label{h1}\end{equation}
where $\delta l$ is the proper distance from $P_{0}$ to 
neighbouring worldlines. This leads to (Ehlers 1993)
\begin{equation}
H_{0}=\left[\frac{1}{3}\Theta+\sigma_{\mu\nu}J^{\mu}
J^{\nu}\right]_{0} \label{hub1}\end{equation}
where 
\begin{equation}
\Theta=u^{\mu}_{\;\; ;\mu}=\frac{\dot{R}^{\prime}}
{R^{\prime}}+2\frac{\dot{R}}{R}\label{hub2}\end{equation}
is the rate of volume expansion, and
\begin{equation}
\sigma_{\mu\nu}=u_{\left(\mu;\nu\right)}-\frac{1}{3}\Theta
\left(g_{\mu\nu}+u_{\mu}u_{\nu}\right)+u_{\left(\mu\right.}
u_{\nu\left.\right);\lambda}u^{\lambda}\label{hub3}
\end{equation}
is the rate of shearing. For the {\sc LTB} metric 
(eq. [\ref{metric}]), this gives
\begin{equation}
\sigma^{1}_{1}=\frac{2}{3}\left(\frac{\dot{R}^{\prime}}
{R^{\prime}}-\frac{\dot{R}}{R}\right),\;\;\;\;\;\;
\sigma^{2}_{2}=\sigma^{3}_{3}=-\frac{1}{2}\sigma^{1}_{1}
\label{hub5}
\end{equation}
with all other $\sigma^{\mu}_{\;\nu}$ vanishing.
Equations (\ref{hub1}-\ref{hub5}), (\ref{j}), and 
(\ref{unangles}) result in
\begin{equation}
H_{0}={\cal C}_{1}+{\cal C}_{2}\cos^{2}\psi
\end{equation}
Definition (\ref{h1}) is equivalent to the (covariant) 
definition of $H_{0}$ as the 
initial slope of the redshift--area distance curve:
\begin{equation}
H_{0}=\left(\frac{dz}{dD}\right)_{0}
\end{equation}
as may be seen from equation (\ref{hresult}).
The deceleration parameter $q_{0}$ may be covariantly
defined as ({\em cf}. Partovi and Mashhoon 1984; 
Kantowski {\em et al}. 1995; Dabrowski 1995)
\begin{equation}
q_{0}=-H_{0}\left(\frac{d^{2}D}{dz^{2}}\right)_{0}-3
\end{equation}
and using equation (\ref{decel}) this results in 
\begin{equation}
q_{0}=-\frac{\C_{3}+
\C_{4}|\cos\psi|+\C_{5}\cos^{2}\psi+\C_{6}|\cos\psi|^{3}}
{\left({\cal C}_{1}+{\cal C}_{2}\cos^{2}\psi
\right)^{2}}-3
\end{equation}
The above formulae for $H_{0}$ and $q_{0}$ are exact covariant 
generalisations of the {\sc FLRW} definitions. 
Note that neither of the expressions possess a dipole.
In principle, they provide an observational test of these models
against the peculiar velocity--type models which do predict 
dipoles. Typical values of the $C_{I}$ for Great--Attractor data are
given in the appendix.

The galactic number count $N$ (per steradian at $P_{0}$) 
is in general dependent on the direction of observation. 
Furthermore, the
total integrated number count over all angles and out to some 
distance, will depend on the distance definition used 
(redshift, area--distance,$\cdots$).
One may obtain $N$ (considered as a function of $\vartheta$ 
and $\varphi$) from\footnote{Equation (19) of Ellis {\em et al}. 
(1985) has an error: the RHS should be multipied by 
${d\upsilon}/{dy}$.} equation (19) of Ellis {\em et al}. (1985):
\begin{equation}
dN=snD^2 (1+z) d\upsilon  \label{N}
\end{equation}
where $n$ is the proper number density of galaxies and $s$ 
is the selection factor (mean fraction of galaxies that are 
detectable). In general $s$ varies
down the null--geodesics; here it is assumed constant since such 
observational problems are not the focus of this paper. The 
function $n$ is determined by $n=\rho/m$, where $m$ is the mean 
galactic mass 
(assumed constant) and $\rho$ is given by equation 
(\ref{density}). 
Equation (\ref{N}) and the local null--geodesic solution lead to
\begin{equation}
\left(\frac{d^{3}N}{dz^{3}}\right)_{0}=\left(\frac{s}{4\pi m}
\right)\frac{\C_{7}}{\left({\cal C}_{1}+{\cal C}_{2}\cos^{2}\psi
\right)^{3}}
\end{equation}
\begin{equation}
\left(\frac{d^{4}N}{dz^{4}}\right)_{0}=\left(\frac{s}{4\pi m}
\right)\frac{\C_{8}+{\cal C}_{9}
|\cos\psi|+\C_{10}\cos^{2}\psi+\C_{11}|\cos\psi|^{3}+\C_{12}
\cos^{4}\psi
}{\left({\cal C}_{1}+{\cal C}_{2}\cos^{2}\psi\right)^{5}} 
\label{endres}
\end{equation}
where the constants $\C_{I}$ are in the appendix. The first two 
derivatives of $N$ with respect to $z$ vanish, by equation 
(\ref{N}). Note that, 
like $H_{0}$ and $q_{0}$, these $N$-$z$ relations do not 
contain any dipoles.

The luminosity distance $D_{L}$ is (Ellis {\em et al}. 1985)
\begin{equation}
D_{L}=D(1+z)^2           \label{lumino}
\end{equation}
All derivatives of $D$, $N$, and $D_{L}$ with respect to $z$ 
at $P_{0}$ 
may be calculated in principle. However, high--order derivatives 
are of no
practical interest as they are well beyond the bounds of current 
observational sensitivity. Given the assumption $z\ll 1$ on the 
boundary $\left\{r=\chi\right\}$, the model is well--characterised 
by ideal observations expanded as far as their first two 
non--vanishing powers in the redshift. 

\section*{5. Off--Center Observations: CMBR}
In this section we present a numerical procedure for calculating 
the {\sc CMBR} anisotropy, and in the appendix the dipole is
evaluated in a special case. 

To obtain the dipole in the 
anisotropy, it is supposed that there exists a 
well--defined last scattering surface from
which the {\sc CMBR} is emitted isotropically, so that we 
exclude the effect on the {\sc CMBR} of processes at decoupling, 
in order to focus on the purely gravitational effect of the 
nonlinear matter inhomogeneity ({\em cf}. Raine and Thomas 1981; 
Tomita and Watanabe 1989; Arnau {\em et al}. 1994). 
We also ignore the 
anisotropies arising from inhomogeneous regions other than our 
own (in the `Swiss cheese' model, see Arnau {\em et al}. (1994) for
detailed calculations). The calculated dipole anisotropy is 
thus due to the off--center location of the observer in the 
inhomogeneous region (figure 1). 
We do not calculate the quadrupole (or higher multipole moments),
as the inhomogeneous--matter contribution to 
the {\sc CMBR} quadrupole is expected to be swamped by effects at 
decoupling (see Nakao {\em et al}. (1995) for
estimates in the case of an observer with peculiar motion).

The dipole anisotropy may be calculated for an off--center 
observer by noting that the first 
two terms of the multipole temperature expansion are $T$ and
$\Delta T\cos\psi$. Calculation of the difference in 
redshift of {\sc CMBR} rays along 
$\left\{\psi=0\right\}$ 
and $\left\{\psi=\pi\right\}$ then suffices to obtain $\Delta T/T$. 
These trajectories are the radial null--geodesics:
\begin{equation}
\frac{dt(r)}{dr}=\left(-1\right)^{\epsilon +1}\frac{R^{\prime}
[r,t(r)]}{\sqrt{1+f(r)}} \label{rad}\end{equation}
With (\ref{rad}), (\ref{z}), and $L\equiv 0$, equation 
(\ref{geo1}) expresses the redshift as 
\begin{equation}
\left[\frac{1}{1+z(r)}\right]\frac{dz(r)}{dr}=(-1)^{\epsilon}
\frac{\dot{R^{\prime}}[r,t(r)]}{\sqrt{1+f(r)}} \label{zode}
\end{equation}
The coupled first order o.d.e.s, (\ref{rad}) and (\ref{zode}), 
are readily solved numerically, with initial data
$\left\{z(r_{0})=0,\;t(r_{0})=t_{0}\right\}$. In the hyperbolic and
elliptic cases (eq. [\ref{f>0 i}--\ref{f<0 i}]) we do not 
have $R[r,t(r)]$
explicitly. For numerical purposes, we write instead $R[r,\eta(r)]$
and solve equations (\ref{rad}--\ref{zode}) coupled with a third 
o.d.e.:
\begin{equation}
\frac{d\eta(r)}{dr}=\dot{\eta}\left[r,\eta(r)\right]
\frac{dt(r)}{dr}+\eta^{\prime}\left[r,t(r),\eta(r)\right]
\end{equation}
The cosmological scenario under consideration in this paper is
illustrated in figure 1. 
Let the proper times at events $A$, $C$,
and $X$ be denoted $t_{A}$, $t_{C}$, and $t_{X}$. The 
redshift between $P_{0}$ and event $X$ (event $C$) is 
$z_{X}$ ($z_{C}$), and 
$z_{AX}$ is that between $A$ and $X$. The {\sc FLRW} 
redshift $z_{BC}$ between $B$ and $C$ is given by
\begin{equation}
1+z_{BC}=\left\{\begin{array}{ll}
\left(t_{C}/t_{B}\right)^{\frac{2}{3}} & (f=0)\\
\left(\cosh\eta_{C}-1\right)/\left(\cosh\eta_{B}-1\right) 
& (f>0)\\
\left(\cos\eta_{C}-1\right)/\left(\cos\eta_{B}-1\right) 
& (0>f>-1)
\end{array}\right\}\equiv\frac{\left({\cal H}^{\frac{2}{3}}
\Omega^{\frac{1}
{3}}\right)_{B}}{\left({\cal H}^{\frac{2}{3}}
\Omega^{\frac{1}{3}}\right)_{C}}\label{flrwz}\end{equation}
and by $t_{B}=t_{A}$. The parameter $\eta$ is defined in 
equations (\ref{flrw1}-\ref{flrw3}). 

Now the redshift in the $\psi=0$ 
direction, minus the redshift in the opposite direction is
\begin{equation}
\Delta z=\left(1+z_{C}\right)\left(1+z_{BC}\right)-\left(1+z_{X}
\right)\left(1+z_{AX}\right)
\label{diform}
\end{equation}
and the exact expression for the temperature dipole anisotropy is
\begin{equation}
\frac{\Delta T}{T}=\frac{1}{2}\left\{ \left(1+z_{C}\right)^{-1}
\left(1+z_{BC}\right)^{-1}
-\left(1+z_{X}\right)^{-1}\left(1+z_{AX}\right)^{-1} \right\}
\label{tempdip}
\end{equation}
which is approximately $-\Delta z/2$ for small redshifts. 

For numerical calculations, the gauge freedom in the radial 
coordinate should be removed. Assuming a finite and non--zero 
proper
density, equation (\ref{feqn}) implies $M(r)=O(r^{3})$ for small 
$r$. Therefore the coordinate freedom in $r$ may be removed by 
choosing a mass--coordinate: 
\begin{equation}
2M(r)=r^{3}   \label{coord}
\end{equation}
such that the remaining (non--trivial) arbitrariness in the 
dust solution (given by two arbitrary functions; Bondi 1947),
corresponds to physically distinct models. 

Equations (\ref{rad}--\ref{coord}) give a simple procedure for 
{\sc CMBR} dipole calculation. A numerical example is given in 
the appendix, reproducing the observed $0.1$\% dipole for a 
reasonable choice of parameters.

\section*{6. Discussion}

In \S 4 we obtained that the angular dependences of $H_{0}$, 
$q_{0}$, $d^{3}N/dz^{3}$, and $d^{4}N/dz^{4}$ 
are functions of $|\cos\psi|$, and hence have no dipole 
components in their multipole expansions. 
Since idealised observations taken further 
than their second non--vanishing redshift derivatives 
($d^{3}D/dz^{3}$, etc.) are well beyond the sensitivity of 
current observational data, it can be concluded that all current
idealised galactic observations should reveal no dipole
dependences, if our model is a valid description of the real
universe. If dipoles were observed, our model would fall away, 
and the peculiar--velocity models would be strengthened.

In the appendix, the main results of this paper are illustrated
by application to the Great--Attractor data. The Hubble parameter 
result
\begin{equation}
H_{0}\approx\left(45+12\cos^{2}\psi\right)\;{\rm kms}^{-1}
{\rm Mpc}^{-1}
\end{equation}
shows a quadrupole term with an amplitude of $\approx 12$\% of the 
magnitude of the monopole. We emphasise 
that the existence of a 
quadrupole component in $H_{0}$ (and in $q_{0}$) has nothing 
in common with `correcting' $H_{0}$ for any `peculiar velocity' 
(such as that defined by the {\sc CMBR} dipole). Furthermore, 
the quadrupole cannot be removed by any such Lorentz 
transformation at $P_{0}$, since the corresponding correction to
$H_{0}$ is dipolar and not quadrupolar (Nakao {\em et al}. 1995). 
The {\sc GA} data, whilst having considerable standard 
deviation, 
do represent a reasonable local geometry, and suggest that 
the angular variations of $H_{0}$ and $q_{0}$ are not 
negligible in many physically acceptable cosmologies.  

The set of physically 
independent arbitrary parameters
at $P_{0}$ is $\{\dot{R},R,R^{\prime},R^{\prime\prime},
f,f^{\prime},f^{\prime\prime}\}$. The mass out to 
$P_{0}$ is given by $\dot{R}$, but higher time derivatives 
are not independent, due to the field equation (\ref{feqn}). 
This set of seven parameters can 
be related to a set of seven independent observed parameters by
an algebraic inversion. For brevity, the result of the calculation 
is not given in this paper. Note however that one 
possible choice of seven independent observed parameters is 
$\left\{\C_{1},\C_{2},\C_{3},\C_{4},\C_{7},\C_{9},
\C_{11}\right\}$.

In {\sc LTB } models, the quadrupole term in $H_{0}$ vanishes 
along off--center comoving
world--lines when ${\cal C}_{2}=0$. This reduces to the pair of
conditions
\begin{equation}
\left(\beta^{\prime}\right)_{0}=0,\;\;\;\;\;\;\left[f^{\prime}
-f\left(4/M\right)^{\frac{1}{3}}\right]_{0}=0 \label{quadconds}
\end{equation}
which imply (in one sense) that the universe is 
locally {\sc FLRW}: the first implies that in the comoving 
frame, the big--bang was locally simultaneous at $r=r_{0}$; the 
second implies that either $f=0$ (corresponding to the 
Einstein--de Sitter model), or that at $r=r_{0}$ the radial 
gradients of the gravitational mass $M$ and total energy $W$
($W\equiv\sqrt{1+f}$, Bondi 1947) are related by
\begin{equation}
\frac{3W}{W^{2}-1}dW=\frac{dM}{M}
\end{equation}
which is satisfied for all $r$ in {\sc FLRW} models.
Hence if the quadrupole vanishes everywhere, the universe must be 
the {\sc FLRW} model.

Wesson (1979) considered an exact self--similar dust solution 
(a special case of {\sc LTB}), and obtained expressions for 
the observables. This paper may be considered a generalisation 
of Wesson (1979), since for the special case he considered, 
\begin{equation}
R(r,t)=r\left[\frac{3}{2}(\alpha_{s}+t/r)\right]^{\frac{2}{3}},
\;\;\;\;\;\;f(r)=0
\end{equation}
his results follow from those here.
Note however that here the full angular dependences of
the observations are given (Wesson considered only the 
azimuthal and radial directions), and that here the luminosity 
distance--redshift relation and number count--redshift relation 
are considered, as opposed to Wesson's (equivalent)
magnitude--redshift and number density--redshift relations. 

While the off--center position of the 
observer removes many of
the simplifications imposed by isotropy, spherical symmetry remains 
a special case of inhomogeneity. 
However, assuming that local inhomogeneities are spherically 
symmetric is a first step towards the study of realistic 
inhomogeneities and their impact on observations. 

\section*{Appendix} 
\subsection*{a. Specific Models using Great--Attractor Data}
The Great--Attractor ({\sc GA}) is generally assumed to cause a
perturbation of the homogeneous background which lies in the linear
regime ({\em cf}. Panek 1992). 
Data for the attractor is typically in the
form of a peculiar velocity field $V$. In particular, Lynden--Bell et
al. (1988) give
\begin{equation}
V=V_{0}S_{0}(S^{-1})_{t=t_{0}} \label{lyn1}
\end{equation}
where $S$ is radial proper distance from the {\sc GA},  
$S_{0}\approx 87\;$Mpc, $V_{0}\approx-570\;$kms$^{-1}$, and we take
${\cal H}_{0}=50\;$kms$^{-1}$Mpc$^{-1}$. Quantitative density 
estimates are far more speculative. 

In the small density contrast and small peculiar gravitational
potential limit of our model, we may identify
\begin{equation}
V=\left(\dot{R}-{\cal H}R\right)_{t=t_{0}} \label{lyn2}
\end{equation}
and $R\approx S$. Note that there is no unique generalisation of 
equation (\ref{lyn2}) to the fully nonlinear case. In our model,
$V$ is not a peculiar velocity, but is used to fix numerical
values in the model.

Hence we require an inhomogeneous {\sc LTB} model which is
specified by equations (\ref{lyn1}) and (\ref{lyn2}), 
and an assumed density
distribution. Note however, that equation (\ref{lyn1}) will be 
simulated only 
approximately, since the matching conditions require 
$V=0$ at $r=\chi$.
Also equation (\ref{lyn1}) is singular at the center, whereas we 
shall require $V=0$ there, on physical grounds.
One possible inhomogeneous model that fits these observational 
constraints is the elliptic solution given by
\begin{eqnarray}
0>&f(r)&=-\alpha r^{2}\left\{1-\lambda_{1}\left[\exp{\left(
-\lambda_{2}r^{2}/\chi^{2}
\right)}-\exp{\left(-\lambda_{2}\right)} \right]\right\} 
\label{elga1}\\
&\beta(r)&=\lambda_{3} \left[\exp{\left(-\lambda_{4}r^{2}/\chi^{2} 
\right)}-\exp{\left(-\lambda_{4}\right)}\right]   \label{elga2}
\end{eqnarray}
which matches to the closed {\sc FLRW} solution (\ref{flrw3})
at $r=\chi$, with $r$ the mass--coordinate [eq. (\ref{coord})]. The 
constants $\lambda_{I}$ are given by
\begin{equation}
\lambda_{1}=80,\;\;\;\lambda_{2}=0.43,
\;\;\;\lambda_{3}\approx 1.96\times 10^{17}{\rm s},\;\;\;
\lambda_{4}=0.25\nonumber 
\end{equation}
and we choose 
$\alpha\approx 1.32\times 10^{-13}${\rm s}$^{-\frac{2}{3}}$, 
so that $\Omega_{0}=1.1$.
The proper radius out to $r=\chi$ is chosen to be 
$\approx 96.9\;$Mpc. 
The ratio of the inhomogeneous proper density to the 
homogeneous background density $\rho/\bar{\rho}$, is plotted 
in figure 2, and figure 3 gives the {\sc CMBR} dipole. 
From the plots, the observed $0.1$\% dipole occurs
at a radius of $\approx 80\;$Mpc, where the proper density is
$\approx 4\times 10^{-30}$gcm$^{-3}$.

The (mathematically simpler) parabolic case is inappropriate
for description of the {\sc GA} (in our model), as it predicts a 
cold spot in the direction of the {\sc GA}. The problem is that the
rate of expansion near the {\sc GA} is greater than in the {\sc FLRW}
region. Interpreted via the peculiar velocity function $V$, this means
$V$ is positive (in the direction {\em away} from the {\sc GA}). 
We can see this quantitatively as
follows. The density ratio is
\begin{equation}
\frac{\rho}{\bar{\rho}}=\frac{t^{2}}{\left(t-\beta
-\frac{2}{3}r\beta^{\prime}\right)\left(t-\beta\right)}
\end{equation}
We must have $\beta^{\prime}(r)\leq 0$ for $r<\chi$, to avoid 
singular
densities for $t-\beta>0$. From (\ref{f=0con}), 
$\beta(\chi)=0$, hence 
the function $\beta$ must be everywhere positive, and the
peculiar velocity
\begin{equation}
V={\cal H}_{0}r\left(\frac{9}{4}\right)^{\frac{1}{3}}
\left(t_{0}-\beta\right)^{-\frac{1}{3}}\beta
\end{equation}
must also be positive.

It is however possible to use the $f=0$ 
assumption to estimate the observational parameters 
(${\cal C}_{I}$), as they are affected only by the local metric,
and there is no evidence to rule out $f\approx 0$ locally to $P_{0}$. 
Hence we take $S=R+\varepsilon$, where $\varepsilon$ 
corrects for the deviation of $S$ from $R$ near the {\sc GA}, 
and is approximately constant on the time slice 
$\left\{t=t_{0}\right\}$ near $P_{0}$. The total 
mass out to distance $S_{0}$ from the {\sc GA} is 
$\approx 5\times 10^{47}$ kg (Panek 1992). This fixes the
mass--coordinate at $P_{0}$, and the observed 
parameters ${\cal C}_{I}$ may be obtained from appendix 
(b) and by 
equating spatial derivatives of equations (\ref{lyn1}) and
(\ref{lyn2}) at $P_{0}$. The results are
\begin{equation}
\left.\begin{array}{l}
\C_{1}=+45 \\ \C_{2}=+12 \end{array}\right\}{\rm kms}^{-1}
{\rm Mpc}^{-1},\;\;\;\;\;\;
\left.\begin{array}{l}
\C_{3}=-0.02 \\ \C_{4}=-0.4 \\
\C_{5}=-0.01 \\ \C_{6}=+0.6 \end{array}\right\}
{\rm kms}^{-1}{\rm Mpc}^{-2}, \nonumber
\end{equation}
\begin{equation}
\C_{7}=+4000\;\;{\rm kgkm}^{-2}{\rm Mpc}^{-1}
,\;\;\;\;\;\;\left.\begin{array}{l}
\C_{8}=-200 \\ \C_{9}=-9000 \\
\C_{10}=-200 \\ \C_{11}=+10^{5} \\
\C_{12}=+5 \end{array}\right\}{\rm kgMpc}^{-3}{\rm km}^{-1}
{\rm s}^{-1}  \nonumber
\end{equation}

\subsection*{b. Coefficients}
The coefficients in equations (\ref{decel}--\ref{endres}) are 
(implicit evaluation at $P_{0}$)
\begin{eqnarray}
&&\C_{1}=\frac{\dot{R}}{R}\nonumber\\
&&\C_{2}=\frac{\dot{R}^{\prime}}{R^{\prime}}-\frac{\dot{R}}
{R}\nonumber\\
&&\C_{3}=\frac{\ddot{R}}{R}-\frac{3\dot{R}^{2}}{R^{2}}
\nonumber\\
&&\C_{4}=3\sqrt{1+f}\left(\frac{\dot{R}}{R^{2}}
-\frac{\dot{R}^{\prime}}
{R^{\prime}R}\right)\nonumber\\
&&\C_{5}=\frac{3\dot{R}^{2}}{R^{2}}-\frac{3\dot{R}^{\prime 2}}
{R^{\prime 2}}-\frac{\ddot{R}}{R}+\frac{\ddot{R}^{\prime}}
{R^{\prime}}
\nonumber\\
&&\C_{6}=\sqrt{1+f}\left(\frac{3\dot{R}^{\prime}}{RR^{\prime}}
+\frac{\dot{R}^{\prime}R^{\prime\prime}}{R^{\prime 3}}
-\frac{3\dot{R}}
{R^{2}}-\frac{\dot{R}^{\prime\prime}}{R^{\prime 2}}\right)
\nonumber\\
&&\C_{7}=\frac{\dot{R}^{2}}{R^{2}}-\frac{f}{R^{2}}
+\frac{2\dot{R}
\dot{R}^{\prime}}{RR^{\prime}}-\frac{f^{\prime}}{RR^{\prime}}
\nonumber\\
&&\C_{8}=-\frac{9\dot{R}^{4}}{R^{4}}+\frac{9\dot{R}^{2}f}{R^{4}}
-\frac{24\dot{R}^{\prime}\dot{R}^{3}}{R^{\prime}R^{3}}
+\frac{12\dot{R}^{2}f^{\prime}}{R^{\prime}R^{3}}+\frac{
6\dot{R}^{\prime}
\ddot{R}\dot{R}}{R^{2}R^{\prime}}-\frac{6\dot{R}^{2}
\ddot{R}^{\prime}}
{R^{2}R^{\prime}}+\frac{6\dot{R}^{2}\dot{R}^{\prime 2}}
{R^{2}R^{\prime 2}}
-\frac{3f^{\prime}\dot{R}^{\prime}\dot{R}}{R^{\prime 2}R^{2}}
-\frac{6f\ddot{R}}{R^{3}}-\frac{6f^{\prime}\ddot{R}}
{R^{\prime}R^{2}}
\nonumber\\
&&\C_{9}=3\sqrt{1+f}\left(-\frac{4f\dot{R}}{R^{4}}
+\frac{6f\dot{R}^{\prime}}
{R^{3}R^{\prime}}+\frac{4\dot{R}^{3}}{R^{4}}
-\frac{f^{\prime\prime}\dot{R}}
{R^{\prime 2}R^{2}}-\frac{10\dot{R}\dot{R}^{\prime 2}}
{R^{2}R^{\prime 2}}-\frac{2\dot{R}^{2}
\dot{R}^{\prime}R^{\prime\prime}}{R^{2}R^{\prime 3}}+\right.
\nonumber\\
&&\;\;\;\left.+\frac{R^{\prime\prime}f^{\prime}\dot{R}}
{R^{2}R^{\prime 3}}+\frac{6f^{\prime}\dot{R}^{\prime}}
{R^{\prime 2}R^{2}}
+\frac{6\dot{R}^{2}\dot{R}^{\prime}}{R^{3}R^{\prime}}
-\frac{6f^{\prime}\dot{R}}{R^{\prime}R^{3}}
+\frac{2\dot{R}^{2}\dot{R}^{\prime\prime}}{R^{2}R^{\prime 2}}
\right)\nonumber\\
&&\C_{10}=-\frac{6f^{\prime}\ddot{R}^{\prime}}{R^{\prime 2}R}
+\frac{15f^{\prime}\dot{R}^{\prime 2}}{R^{\prime 3}R}
-\frac{12\dot{R}f\dot{R}^{\prime}}{R^{3}R^{\prime}}
+\frac{18f\dot{R}^{\prime 2}}{R^{2}R^{\prime 2}}
-\frac{6f\ddot{R}^{\prime}}{R^{2}R^{\prime}}
-\frac{6\dot{R}^{\prime 2}\ddot{R}}{RR^{\prime 2}}
-\frac{30\dot{R}\dot{R}^{\prime 3}}{RR^{\prime 3}}
+\frac{6\dot{R}\ddot{R}^{\prime}\dot{R}^{\prime}}{RR^{\prime 2}}
-\frac{6\dot{R}^{2}f}{R^{4}}+\nonumber\\
&&\;\;\;-\frac{6\dot{R}^{2}\dot{R}^{\prime 2}}{R^{2}R^{\prime 2}}
-\frac{6f^{\prime}\dot{R}^{\prime}\dot{R}}{R^{2}R^{\prime 2}}
+\frac{30\dot{R}^{\prime}\dot{R}^{3}}{R^{\prime}R^{3}}
-\frac{9\dot{R}^{2}f^{\prime}}{R^{3}R^{\prime}}
-\frac{12\dot{R}^{\prime}\ddot{R}\dot{R}}{R^{2}R^{\prime}}
+\frac{12\dot{R}^{2}\ddot{R}^{\prime}}{R^{2}R^{\prime}}
+\frac{6f\ddot{R}}{R^{3}}+\frac{6f^{\prime}\ddot{R}}
{R^{2}R^{\prime}}
+\frac{6\dot{R}^{4}}{R^{4}}\nonumber\\
&&\C_{11}=-3\sqrt{1+f}\left(\frac{4f\dot{R}^{\prime}}
{R^{3}R^{\prime}}
-\frac{4f\dot{R}}{R^{4}}+\frac{2f\dot{R}^{\prime}
R^{\prime\prime}}
{R^{\prime 3}R^{2}}-\frac{2f\dot{R}^{\prime\prime}}
{R^{\prime 2}R^{2}}
-\frac{2\dot{R}\dot{R}^{\prime 2}R^{\prime\prime}}
{RR^{\prime 4}}
-\frac{f^{\prime\prime}\dot{R}}{R^{2}R^{\prime 2}}
+\frac{6f^{\prime}\dot{R}^{\prime}}{R^{2}R^{\prime 2}}
-\frac{2\dot{R}^{\prime 3}}{RR^{\prime 3}}
-\frac{2f^{\prime}\dot{R}^{\prime\prime}}{RR^{\prime 3}}
+\right.\nonumber\\
&&\;\;\;\left.+\frac{f^{\prime\prime}\dot{R}^{\prime}}
{RR^{\prime 3}}
+\frac{R^{\prime\prime}f^{\prime}\dot{R}}{R^{2}R^{\prime 3}}
+\frac{2\dot{R}\dot{R}^{\prime}\dot{R}^{\prime\prime}}
{RR^{\prime 3}}
+\frac{4\dot{R}^{2}\dot{R}^{\prime\prime}}{R^{2}R^{\prime 2}}
-\frac{6f^{\prime}\dot{R}}{R^{\prime}R^{3}}
+\frac{4\dot{R}^{3}}{R^{4}}+\frac{f^{\prime}\dot{R}^{\prime}
R^{\prime\prime}}
{RR^{\prime 4}}-\frac{10\dot{R}\dot{R}^{\prime 2}}
{R^{2}R^{\prime 2}}
-\frac{4\dot{R}^{2}\dot{R}^{\prime}R^{\prime\prime}}
{R^{\prime 3}R^{2}}
+\frac{8\dot{R}^{2}\dot{R}^{\prime}}{R^{3}R^{\prime}}
\right)\nonumber\\
&&\C_{12}=\frac{3\dot{R}^{4}}{R^{4}}-\frac{3\dot{R}^{2}f}
{R^{4}}+\frac{6\dot{R}\dot{R}^{\prime 3}}{RR^{\prime 3}}
-\frac{3f^{\prime}\dot{R}^{\prime 2}}{R^{\prime 3}R}
-\frac{9\dot{R}^{2}\dot{R}^{\prime 2}}{R^{\prime 2}R^{2}}
-\frac{3\dot{R}^{2}f^{\prime}}{R^{3}R^{\prime}}
-\frac{3f\dot{R}^{\prime 2}}{R^{2}R^{\prime 2}}
+\frac{6\dot{R}f\dot{R}^{\prime}}{R^{3}R^{\prime}}
+\frac{6\dot{R}^{\prime}\dot{R}f^{\prime}}{R^{2}R^{\prime 2}}
\nonumber\end{eqnarray}

\section*{References}
\begin{description}
\item Arnau, J. V., Fullana M. J., and S\'{a}ez, D., 1994, 
{\em M. N. R. A. S.}, {\bf 268}, L17.
\item Bertschinger, E., and Dekel, A., 1989, {\em Ap. J.}, 
{\bf 336}, L5.
\item Bertschinger, E., and Juszkiewicz, R., 1988, {\em Ap. J.}, 
{\bf 334}, L59.
\item Bildhauer, S., and Futamase, T., 1991, {\em M. N. R. A. S.}, 
{\bf 249}, 126.
\item Bondi, H., 1947, {\em M.N.R.A.S.}, {\bf 107}, 410.  
\item Dabrowski, M. P., 1995, {\em Ap. J.}, {\bf 447}, 43. 
\item Ehlers, J., 1993, {\em Gen. Rel. Grav.}, {\bf 25}, 1225 
(translation of original 1961 article).
\item Ellis, G. F. R., Nel, S. D., Maartens, R., Stoeger, W. R., 
and Whitman, A. P., 1985, {\em Phys. Rep.}, {\bf 124}, 315.
\item Israel, W., 1966, {\em Nuovo Cimento}, {\bf 44B}, 1.   
\item Kantowski, R., Vaughan, T., and Branch, D., 1995, 
{\em Ap. J.}, {\bf 447}, 35.
\item Langlois, D., and Piran, T., 1995, preprint astro-ph. 
\item Lynden-Bell, D., Faber, S. M., Burstein, D., Davies, 
R. L., Dressler, A., Terlevich, R. J., and Wegner, G., 1988, 
{\em Ap. J.}, {\bf 326}, 19.  
\item Maartens, R., Humphreys, N. P., Matravers, D. R., 
and Stoeger, W. R., 1996, {\em Class. Quant. Grav.}, in press.
\item M\'{e}sz\'{a}ros, A., 1994, {\em Ap. J.}, {\bf 423}, 19.  
\item Moffat, J. W., and Tatarski, D. C., 1995, {\em Ap. J.}, 
{\bf 453}, 17.
\item Nakao, K., Gouda, N., Chiba, T., Ikeuchi, S., Nakamura, T., 
and Shibata, M., 1995, {\em Ap. J.}, {\bf 453}, 541.
\item Paczynski, B., and Piran, T., 1990, {\em Ap. J.}, 
{\bf 364}, 341.
\item Panek, M., 1992, {\em Ap. J.}, {\bf 388}, 225. 
\item Papapetrou, A., 1978, Ann. L'Institut Henri Poincar\'{e},
{\underline A}, XXIX, 2, 207.
\item Partovi, M. H., and Mashhoon, B., 1984, {\em Ap. J.}, 
{\bf 276}, 4. 
\item Raine, D. J., Thomas, E. G., 1981, {\em M. N. R. A. S.},
{\bf 195}, 649.
\item Riess, A. G., Press, W. H., and Kirshner, R. P., 1995, 
{\em Ap. J.}, {\bf 438}, L17.
\item Tomita, K., 1995, {\em Ap. J.}, {\bf 451}, 1. 
\item Tomita, K., and Watanabe, K., 1989, {\em Prog. Theor. Phys.}, 
{\bf 82}, 563.
\item Van den Bergh, S., 1995, {\em Ap. J.}, {\bf 453}, L55.
\item Wesson, P. S., 1979, {\em Ap. J.}, {\bf 228}, 647.
\end{description}
\section*{Figure Captions}
\[ \]
{\bf Figure 1. }Schematic for the dipole anisotropy. The 
frequencies of the two {\sc CMBR} rays are equal at $A$ and $B$, 
but are then redshifted differently on arrival at $P_{0}$. The 
dotted lines denote constant--$t$ hypersurfaces.
\[ \]
\[ \]
{\bf Figure 2. }Plot of the density ratio $\rho/\bar{\rho}$ 
against radial proper distance $S$ for the elliptic {\sc GA} 
model (eq. [\ref{elga1}--\ref{elga2}]) at time $t=t_{0}$.
\[ \]
\[ \]
{\bf Figure 3. }Plot of the {\sc CMBR} temperature dipole 
$\Delta T/T$ against radial proper distance $S$ for the 
elliptic {\sc GA} model (eq.[\ref{elga1}--\ref{elga2}]) 
at time $t=t_{0}$. 
\[ \]
\[ \]
\begin{figure}[h]
\setlength{\unitlength}{1mm}
\begin{picture}(50,50)(0,0)
\put(53,50){\scriptsize FIGURE 1}
\put(20,0){\framebox(80,45){}}
\put(40,5){\framebox(0,35){}}
\put(70,5){\framebox(0,35){}}
\put(70,29){$\hookleftarrow$}
\put(72,32){\raisebox{-.4ex}{$r=\chi$}}
\put(29,32){$r=\chi$}
\put(35,29){$\hookrightarrow$}
\put(60,5){\vector(0,1){35}}
\put(55,5){\vector(0,2){35}}
\put(30,5){\vector(1,1){15}}
\put(90,5){\vector(-1,1){15}}
\put(45,20){\line(1,1){15}}
\put(75,20){\line(-1,1){15}}
\put(60,35){$P_{0}$} 
\put(44,38){$r=0$}
\put(50,36){\raisebox{0ex}{$\hookrightarrow$}}
\put(52,30){X}
\put(37,15){\raisebox{.4ex}{A}} \put(80,15){\raisebox{.4ex}{B}}
\put(70,25){\raisebox{-.1ex}{C}}
\put(25,15){\dashbox(70,0){}}
\put(75,10){\scriptsize CMBR}
\put(25,5){\dashbox(70,0){}}
\put(25,35){\dashbox(70,0){}}
\put(45,0){\scriptsize\raisebox{1.5ex}{last scattering surface}} 
\end{picture}
\end{figure}
\end{document}